# Double-Crucible Vertical Bridgman Technique for Stoichiometry-Controlled Chalcogenide Crystal Growth

Yingdong Guan[1], Suguru Yoshida[2], Jairo Obando-Guevara[1], Seng Huat Lee[1,2], Heike Pfau[1], and Zhiqiang Mao[1,2*]

[1]*Department of Physics, The Pennsylvania State University, University Park, Pennsylvania 16802, USA*

[2]*2D Crystal Consortium, Materials Research Institute, The Pennsylvania State University, University Park, Pennsylvania 16802, USA*

## Abstract

**Precise stoichiometry control in single-crystal growth is essential for both technological applications and fundamental research. However, conventional growth methods often face challenges such as non-stoichiometry, compositional gradients, and phase impurities, particularly in non-congruent melting systems. Even in congruent melting systems like $Bi_2Se_3$, deviations from the ideal stoichiometric composition can lead to significant property degradation, such as excessive bulk conductivity, which limits its topological applications. In this study, we introduce the double-crucible vertical Bridgman (DCVB) method, a novel approach that enhances stoichiometry control through the combined use of continuous source material feeding, traveling-solvent growth, and liquid encapsulation, which suppresses volatile element loss under high pressure. Using $Bi_2Se_3$ as a model system, we demonstrate that crystals grown via DCVB exhibit enhanced stoichiometric control, significantly reducing defect density and achieving much lower carrier concentrations compared to those produced by conventional Bridgman techniques. Moreover, the continuous feeding of source material enables the growth of large crystals. This approach presents a promising strategy for synthesizing high-quality, large-scale crystals, particularly for metal chalcogenides and pnictides that exhibit challenging non-congruent melting behaviors.**

*Email: zim1@psu.edu

## 1. INTRODUCTION

Large single crystals with precise stoichiometry control are the cornerstones of modern-day microelectronics, global positioning systems, optical communications, energy harvesting, etc. In addition, fundamental research also relies heavily on ideal single crystals to access intrinsic properties (including emergent quantum phases) and discover value in new material space. A grand challenge in single-crystal growth is that high-quality crystals at practical scales can be grown only for congruently melting compositions. Crystal growth from the melt and solution generally encounters four distinct scenarios: (a) congruent melting, (b) solid solution, (c) incongruent melting, and (d) eutectic systems [1]. Congruent compositions are the only ones where the first solid formed upon cooling has nominally the same composition as the liquid from which it precipitates. This is the desired crystallization pathway and is unfortunately not common.

In contrast, each of the alternative scenarios presents challenges such as phase impurities, structure inhomogeneities or composition gradients. Single crystals of these systems can be grown using conventional Czochralski (CZ) [(a), (b)], Bridgman [(a), (b)], top seeding solution growth (TSSG) [(c), (d)] and traveling heater methods [(c), (d)]. However, these approaches often come with significant drawbacks, such as non-stoichiometry and compositional gradients along the growth axis. Despite these challenges, there are numerous instances where single crystals from non-ideal systems exhibit compelling properties. Nevertheless, issues with reproducibility, scale, and defect chemistry often restrict these materials to academic research rather than commercial applications or prevent the attainment of the desired properties.

Even for congruent systems, improvements in crystal growth are badly needed because while close, the congruently solidified crystal composition is usually not precisely stoichiometric, and this discrepancy can cause dramatic property degradation. This situation is exemplified by $Bi_2Se_3$, a prototype topological insulator [2-6]. Although this material is predicted to be promising for developing energy-efficient electronics due to its topological surface states that can support dissipationless currents in theory, its practical applications are hindered by bulk conductivity. This issue arises from heavy electron-doping caused by the non-stoichiometric composition of the crystals, which interferes with surface conduction and limits the performance of topological insulators.

Here, we introduce a novel Bridgman-type growth technique, the double-crucible vertical Bridgman (DCVB) method which can control crystal stoichiometry. Unlike conventional Bridgman growth, where crystals are grown from the melt in a single crucible moving from the hot to cold zone (Fig. 1a), the DCVB furnace employs a two-crucible geometry, as shown in Fig. 1b. This design allows for the continuous feeding of source material from an upper crucible to the growth crucible, facilitating traveling solvent growth within the Bridgman furnace. Combined with liquid encapsulation and high pressure, both of which effectively suppress the vaporization of volatile elements, this approach allows for better control of melt composition throughout the growth process. As a result, it can enable crystal growth for incongruent melting compositions and yield crystals with improved homogeneity and stoichiometry. In this study, we use $Bi_2Se_3$ as an example to demonstrate the capability of the DCVB method. We show that

Bi$_2$Se$_3$ crystals grown with this technique exhibit significantly lower carrier density and greater compositional homogeneity compared to those grown by the conventional Bridgman method.

## 2. CONCEPT OF THE DOUBLE-CRUCILBE APPROACH

The double-crucible technique has a long history and was first successfully implemented with the Czochralski method for growing oxide single crystals [7,8]. A notable example of stoichiometric crystal growth using the double-crucible Czochralski method is LiNbO$_3$, a cornerstone non-linear optical material that underpins the electro-optics that power the internet [9,10]. Its congruent-melting composition is nonstoichiometric, Li$_{0.94}$NbO$_3$, containing Li vacancies (V$_{Li}$) and Nb$_{Li}$ antisite defects. Stoichiometric LiNbO$_3$ is in equilibrium with a Li-rich liquid phase (8–10 mol% excess Li), allowing the growth of stoichiometric LiNbO$_3$ from such a melt. However, as the growth proceeds, the melt composition changes over time, leading to compositional gradients in the final crystal. By continuously supplying stoichiometric melt from the outer crucible, the melt in the inner crucible remains Li-rich, enabling the growth of stoichiometric LiNbO$_3$ throughout the entire process [7].

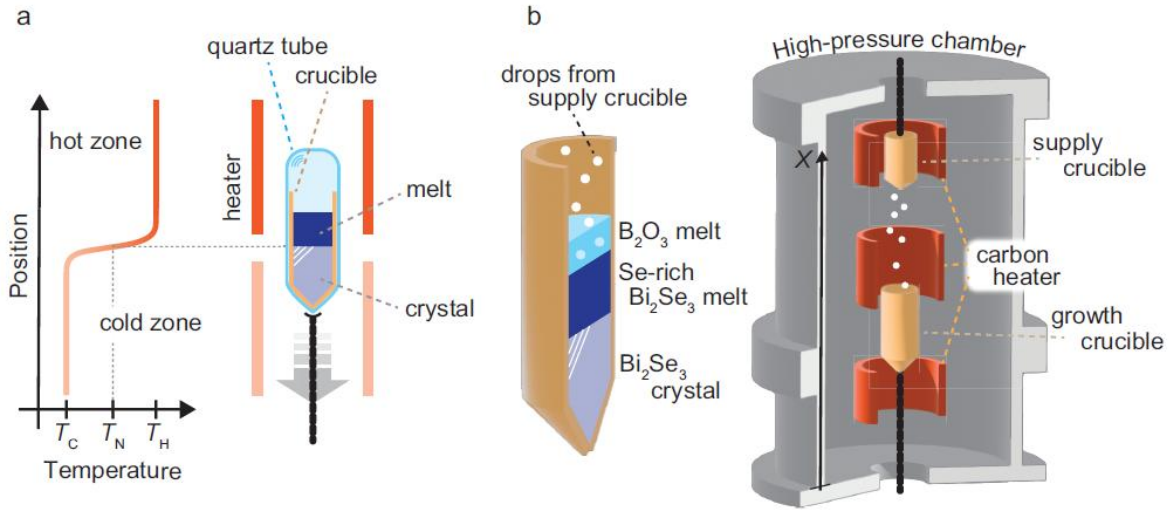

FIG. 1: Schematic description of (a) conventional Bridgman growth and (b) DCVB technique developed through this work. $T_H$, $T_C$, and $T_N$ represent the hot zone, cold zone, and nucleation temperatures of the targeted crystal, respectively. In the conventional Bridgman process, the starting material is often charged directly in the quartz tube when it makes no reaction with SiO$_2$. The left part of panel (b) shows the cross-sectional illustration of the growth-crucible configuration used for the DCVB growth of Bi$_2$Se$_3$.

The DCVB furnace adapts the double-crucible concept to the Bridgman technique and is specifically designed for growing chalcogenide/pnictide single crystals with controlled compositions. The furnace is equipped with two crucibles serving distinct functions, as well as a large high-pressure chamber (Fig. 1b), into which the starting materials are loaded without being sealed in an ampule. The upper crucible (supply crucible) continuously feeds source material into the lower crucible (growth crucible), where the crystal grows. This unique design allows for the maintenance of melt composition throughout the growth process, enabling the production of large, spatially homogeneous crystals even when the target composition exhibits incongruent-melting behavior.

The furnace is also designed to apply high pressure, using inert gas (up to 10 atm), which suppresses the vaporization of volatile elements such as chalcogens and further enhances the crystalline quality. Additionally, the DCVB method incorporates another technique to minimize material loss due to vaporization: liquid encapsulation. This concept, originally developed for the Czochralski process [11-13], was later adapted for use with Bridgman techniques [14-16]. In this method, high-purity $B_2O_3$ powder is placed in the growth crucible on top of the raw material. As the temperature increases, $B_2O_3$ (with a melting point of approximately 450°C) melts before the raw material does, forming a protective liquid layer over the melt zone. The $B_2O_3$ layer acts as a lid, effectively minimizing the vaporization of volatile components from the melt zone, while still allowing material from the supply crucible to pass through and merge with the melt (left panel of Fig. 1b). Other materials can also be used as encapsulants, provided they are chemically inert and have lower melting points and densities than those of the melt zone.

### 3. GROWTH OF $Bi_2Se_3$ CRYSTALS USING THE DCVB TECHNIQUE

**DCVB growth mechanism for $Bi_2Se_3$**

The DCVB facility at Penn State, manufactured by Oxide Corporation, is the first of its kind in academia. We have validated its growth capabilities by successfully synthesizing low carrier density $Bi_2Se_3$. While stoichiometric $Bi_2Se_3$ is theoretically predicted to be an insulator with a gap of 0.3 eV and host an in-gap topological surface state, experimentally grown $Bi_2Se_3$ crystals - produced via melt and conventional vertical Bridgman (CVB) methods - are either heavily or moderately electron-doped, with carrier densities ranging from $10^{18}$ to $10^{19}$ cm$^{-3}$ for CVB grown crystals (see Table 1). This excessive electron doping primarily arises from crystal defects due to its nonstoichiometric congruent melting composition, including Se vacancies ($V_{Se}$), interstitial Se ($Se_i$), and antisite defects ($Bi_{Se}$) [17]. Although the Se-flux growth method can mitigate these defects and reduce the carrier density to $10^{16}$-$10^{17}$ cm$^{-3}$ [18-21], achieving large-scale, homogeneous crystals with such low doping levels remains challenging. This difficulty arises because a high Se concentration causes the source material to have an incongruent melting composition closer to the eutectic point (see the phase diagram in Supplementary Fig. S1), ultimately leading to the formation of a Se-$Bi_2Se_3$ mixture in the final product. Although excess Se flux can be removed by centrifugation or decanting, obtaining large-sized crystals suitable for practical applications remains challenging. Several groups have also attempted Se-rich flux growth in a vertical Bridgman furnace [22,23]. While this method can reduce the carrier density of the grown crystals to the order of $10^{17}$ cm$^{-3}$, the resulting crystals are mixed with Se precipitates. In the following section, we demonstrate how the DCVB technique effectively addresses this challenge.

The significantly lower carrier densities observed in $Bi_2Se_3$ single crystals grown with rich Se flux [18-21] suggest that stoichiometric $Bi_2Se_3$ is in equilibrium with a Se-rich phase, as indicated by the Bi-Se phase diagram in Supplementary Fig. S1. This implies that stoichiometric $Bi_2Se_3$ crystals can probably be grown from incongruent melting compositions containing excess Se. To prevent the formation of a Se-$Bi_2Se_3$ mixture in the final product, the traveling solvent growth concept must be implemented. Specifically, stoichiometric $Bi_2Se_3$ source material should be continuously supplied to the traveling melt, where excess Se in the melt acts as the solvent. The melt composition remains stable when the source material feeding rate closely matches the crystal growth rate. Crystals grown under these conditions are expected to be homogeneous and stoichiometric or close to be stoichiometric.

The DCVB furnace is specifically designed to enable such a traveling solvent growth within a Bridgman setup. This method bears similarities to the traveling solvent floating-zone (TSFZ) technique, where source material is continuously fed into a flux-containing melting zone while the seed crystal moves downward to sustain growth. TSFZ is highly effective for growing incongruently melting oxides but is not suitable for chalcogenides due to the high volatility of chalcogen elements. However, as discussed above, the combination of liquid encapsulation and high-pressure conditions in the DCVB furnace can significantly suppress chalcogen vaporization. Consequently, the DCVB technique is expected to be highly effective for the growth of chalcogenide materials via the traveling solvent method with continuous source feeding.

To investigate this novel traveling solvent Bridgman growth mechanism, we synthesized $Bi_2Se_3$ single crystals in the DCVB furnace both with and without the feeding function. The batch grown without source material feeding is designated as DCVB1 below, while the batch grown with continuous source material feeding is labeled DCVB2. Comparing these two batches allows us to evaluate the crucial role of continuous source material feeding in achieving low-carrier-density crystal growth. As described in the Methods section, the source material initially loaded into the growth crucible had a Bi:Se molar ratio of 34:66 (~2:3.88), corresponding to a Se-rich composition, with excess Se (29.3%) serving as a flux. The detailed growth procedures for both batches are provided in the Methods section.

**DCVB growth without source material feeding (DCVB1)**

Figure 2(a) shows an image of the final DCVB1 product extracted from the alumina growth crucible. The top yellow layer is composed of $B_2O_3$, which acted as a liquid encapsulant during the crystal growth process. Although $B_2O_3$ is relatively chemically inert, it reacts with quartz; therefore, an alumina crucible was chosen for $Bi_2Se_3$ crystal growth. From this growth batch, we also observed that the $B_2O_3$ liquid encapsulation, in combination with high pressure, effectively minimized Se vaporization from the growth crucible. This is supported by the minimal Se contamination observed on the growth chamber wall.

To examine the structural phases and composition of different regions of the crystal ingot, we performed X-ray diffraction (XRD) and energy-dispersive X-ray spectroscopy (EDXS) measurements. Based on these analyses, the ingot was divided into three distinct regions: A, B, and C (Fig. 2a). Part A, located directly beneath the $B_2O_3$ layer, is primarily composed of Se with a minor $Bi_2Se_3$ phase. No evidence of $B_2O_3$ reacting with Se or $Bi_2Se_3$ was observed, confirming that $B_2O_3$ is an effective encapsulant for $Bi_2Se_3$ growth. Part B consists of a mixture of $Bi_2Se_3$ and Se, as evident from the composition mapping of a cleaved crystal sample (Fig. 2b), where the green regions correspond to $Bi_2Se_3$ and the red regions to pure Se. Pure $Bi_2Se_3$ crystals were obtained only from the bottom region of the ingot (Part C), which had a significantly smaller volume compared to Parts A and B, indicating a low crystal yield.

To further investigate compositional homogeneity, we performed EDXS line scans on multiple crystal pieces, each a few millimeters in lateral dimensions, cleaved from Part C. While some samples exhibited uniform composition, others displayed noticeable variations. A representative scan along Line 1 in Fig. 2c (top panel, Fig. 2d) shows negligible compositional fluctuation. In contrast, the scan along Line 2 in Fig. 2c (bottom panel, Fig. 2d) reveals significant variations, with the Se concentration ranging from 60% to 57.5% and Bi from 40% to 42.5%.

These findings clearly indicate that simply combining Bridgman growth with excessive Se-flux growth is not an effective method for producing large, homogeneous Bi₂Se₃ crystals, consistent with prior reports [22,23]. More than 50% of the source material did not form pure Bi₂Se₃ but instead crystallized as a mixture of Bi₂Se₃ and Se. This result aligns with the combined Bridgman-flux growth mechanism discussed earlier and can be understood as follows: during crystal growth, the melt initially contained 29.3% excess Se as a flux. As Bi₂Se₃ crystallized from the melt, the relative proportion of Se flux increased, shifting the melt composition toward the Bi₂Se₃-Se eutectic point. Consequently, a significant fraction of the ingot solidified as a Bi₂Se₃-Se mixture, explaining why the volume of the mixed-phase region (Part B) exceeds that of the pure Bi₂Se₃ region (Part C).

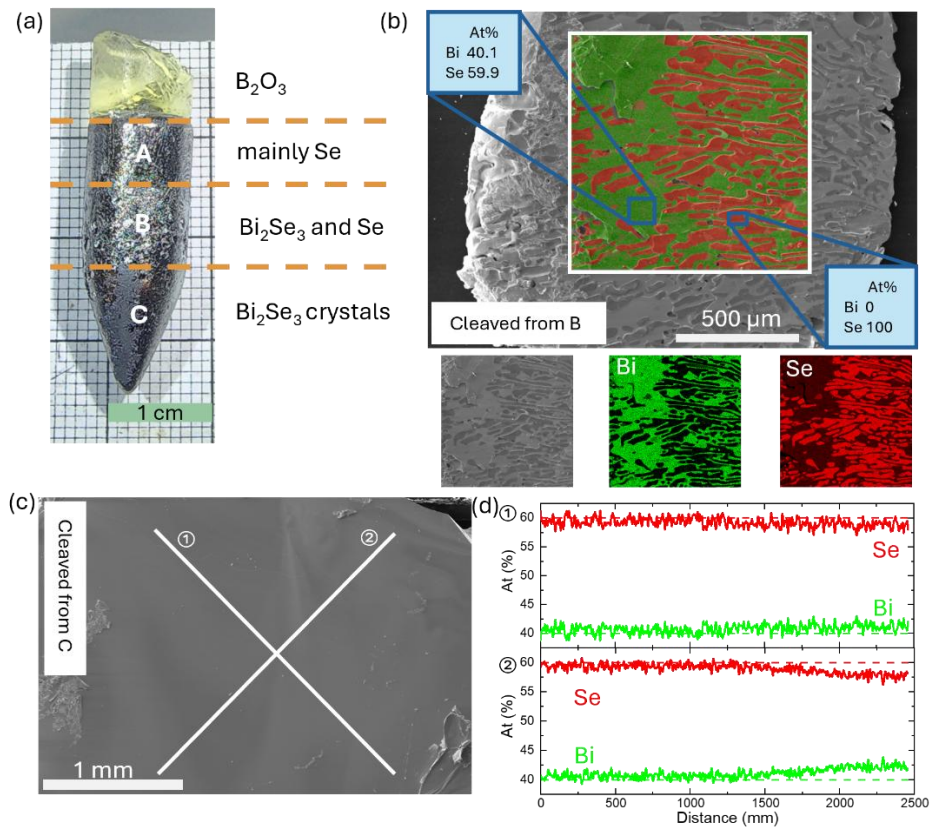

Figure 2. (a) Image of the crystal ingot grown using the DCVB furnace without feeding. The crystal can be divided into four regions: B₂O₃, a predominantly Se region (A), a Bi₂Se₃-Se mixture region (B), and a pure Bi₂Se₃ single-crystal region (C). (b) EDXS mapping of a single crystal cleaved from region B. (c) Scanning electron microscopy (SEM) image and (d) line EDXS scan of a single crystal cleaved from region C, where ① and ② in (c) and (d) indicate the scanned lines.

**DCVB growth with source material feeding (DCVB2)**

The issues of low crystal yield and inhomogeneous composition encountered in the DCVB1 growth were effectively resolved in the DCVB growth with continuous source material feeding (DCVB2), as expected. Figure 3a shows the crystal ingot obtained from the DCVB2 growth process. The top gray layer consists of $Bi_2O_3$, which appears gray rather than yellow (Fig. 2a) due to the incorporation of a small amount of $Bi_2Se_3$ within $B_2O_3$ as it solidified during cooling. This occurred because the $Bi_2Se_3$ source material was fed into the growth crucible through the liquid $B_2O_3$ encapsulation layer.

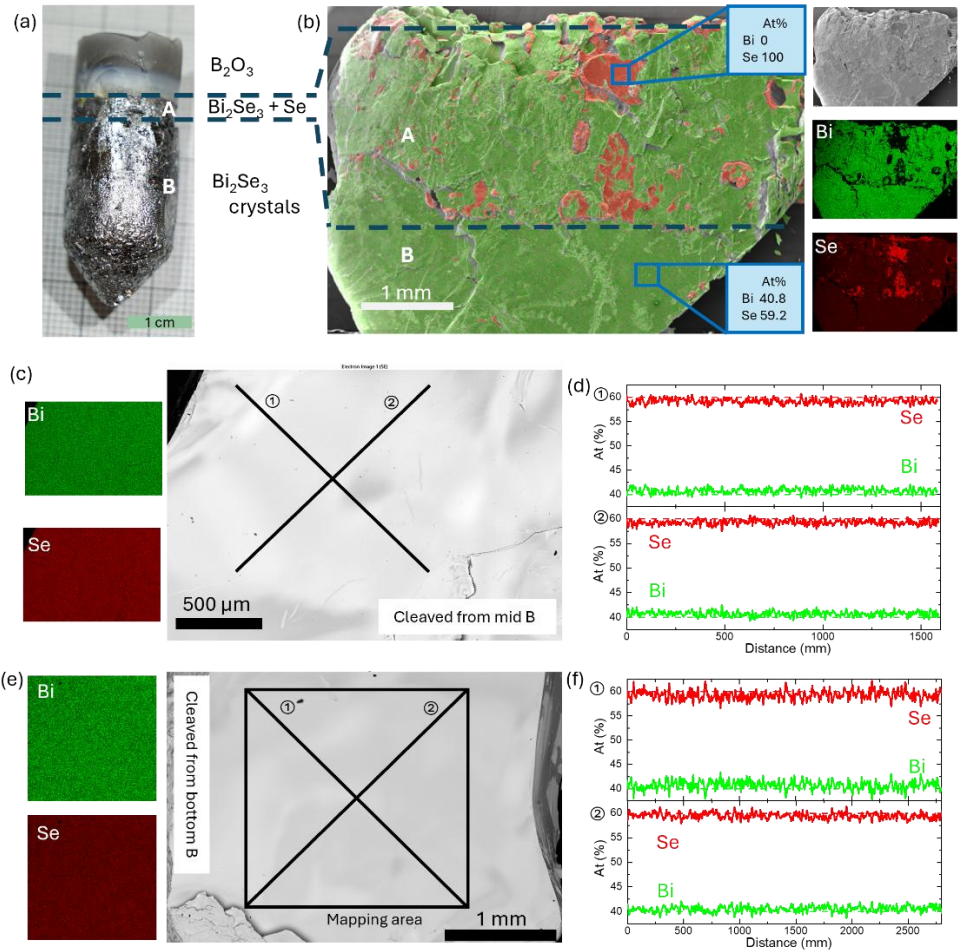

Figure 3. (a) Image of the crystal grown using the DCVB furnace with continuous feeding (DCVB2). The crystal is divided into three regions: $B_2O_3$, a $Bi_2Se_3$-Se mixture region (A), and a pure $Bi_2Se_3$ single-crystal region (B). (b) SEM-EDXS mapping of a single crystal cleaved from the top part of the ingot, containing material from both regions A and B. (c, e) SEM-EDXS mapping of single crystals cleaved from the middle and bottom of region B, respectively. (d, f) SEM-EDS line scans for single crystals cleaved from the middle and bottom of region B, respectively, where ① and ② in (d) and (f) indicate the scanned lines in (c) and (e).

To analyze the crystal structure and composition distribution in the DCVB2 ingot, we performed XRD and EDXS on crystals cleaved from various regions. The composition distribution in DCVB2 differs significantly from that of DCVB1. First, no region analogous to Part A in DCVB1, which was predominantly Se, was observed. Second, while DCVB2 also contains a section with a Bi$_2$Se$_3$-Se mixture (Part A in Fig. 3a), its volume fraction is considerably smaller than that of the mixed-phase region in DCVB1 (Part B in Fig. 2a). As shown in Fig. 3b, Bi$_2$Se$_3$ is the dominant phase in this section. Third, due to continuous source material feeding, the pure Bi$_2$Se$_3$ region (Part B in Fig. 3a) is significantly larger than the corresponding region in DCVB1 (Part C in Fig. 2a). This demonstrates that traveling-solvent growth combined with continuous source material feeding greatly enhances crystal yield, enabling the growth of large-dimension Bi$_2$Se$_3$ crystals. With larger-diameter crucibles, Bi$_2$Se$_3$ crystals exceeding an inch in diameter can potentially be grown.

Although the Bi$_2$Se$_3$ crystals in Part B exhibit multiple domains, EDXS composition analysis indicates that most cleaved crystals from this region exhibit a highly homogeneous composition. Figures 3c–3f present two representative examples. Figures 3c–3d show the results for a crystal cleaved from the middle of Part B in Fig. 3a. The cleaved surface is notably flat (Fig. 3c, right panel), and composition mapping (Fig. 3c, left panel) confirms a uniform composition. This homogeneity is further supported by EDXS line scans, which show no observable variation in the Bi and Se atomic ratio over a ~1.5 mm length scale along the two scanned lines marked in Fig. 3c (right panel). The atomic ratio of Bi and Se remains consistently at about 40% and 60%, respectively (Fig. 3d). Similar results were obtained for a crystal cleaved near the bottom of Part B in Fig. 3a (Figs. 3e–3f), where no composition variation was observed even over a ~3 mm length scale.

The difference in crystal yield and compositional homogeneity between DCVB1 and DCVB2 underscores the effectiveness of traveling solvent growth in the DCVB furnace for producing large Bi$_2$Se$_3$ crystals with a relatively uniform composition. The relatively small volume fraction of mixed Bi$_2$Se$_3$ and Se can be explained as follows. Although the initial molar ratio of Bi:Se in the source material was the same (34:66) for both methods, the total mass of material initially loaded into the growth crucible differed: 17.2 g for DCVB2, nearly twice the 9.8 g used for DCVB1. This difference arose because DCVB2 required a significantly larger crucible to accommodate continuous source material feeding. The total amount of Bi$_2$Se$_3$ supplied from the feeding crucible to the growth crucible during DCVB2 growth was approximately 16 g. As the Bi$_2$Se$_3$ melt in the supply crucible gradually dripped into the growth crucible, chemical vapor of Se, Bi and Bi$_2$Se$_3$ formed and deposited on the chamber walls, as evidenced by contamination observed in the growth chamber. Given that Se has a much higher vapor pressure, the composition of the melt dripping into the growth crucible should be Se-deficient. However, the excess Se flux in the growth crucible compensated for this Se loss, explaining why significantly less residual Se was observed in DCVB2 compared to DCVB1 (Part A, Fig. 3a). This also suggests that near the end of growth, the excessive Se flux in the growth crucible may be insufficient to fully counteract Se loss. Consequently, the melt composition likely drifted toward the congruent melting composition, leading to less mixed Bi$_2$Se$_3$ and Se phase in part A and inhomogeneous crystal composition near the upper section of Part B in Fig. 3a. Indeed, transport and band structure measurements, discussed below, reveal characteristics consistent with this composition inhomogeneity.

**Electronic properties and band structure of DCVB1 and DCVB2 crystals**

Although Bi$_2$Se$_3$ exhibits several types of defects, as noted above, its primary defect is Se vacancies [17], which act as donors and contribute to electron doping. Most previously reported Bi$_2$Se$_3$ crystals grown using conventional vertical Bridgman (CVB) furnaces exhibit moderate to heavy electron doping, with carrier densities ranging from $2.0\times10^{18}$ to $5.6\times10^{20}$ cm$^{-3}$ (see Table 1). Only those crystals mixed with Se precipitates exhibit a carrier density reduced to the order of $10^{17}$ cm$^{-3}$ [22]. For comparison with DCVB growth, we also synthesized Bi$_2$Se$_3$ crystals using the CVB method with 2% excessive Se flux. As shown in Table 1, our CVB-grown Bi$_2$Se$_3$ crystals have carrier densities in the range of $1.3\times10^{19}$ - $1.6\times10^{19}$ cm$^{-3}$, consistent with previously reported values. To determine whether Bi$_2$Se$_3$ crystals grown using the DCVB method exhibit a reduced density of Se vacancies as expected, we performed electronic transport measurements on both DCVB1 and DCVB2 crystals. Carrier densities were evaluated at 6 K using field-dependent Hall resistivity measurements on multiple crystal pieces cleaved from different regions of part C of DCVB1 and part B of DCVB2 ingots (see Fig. 4b for representative Hall resistivity data). The results show that pure Bi$_2$Se$_3$ crystals from the DCVB1 batch have carrier densities ranging from $1.9\times10^{17}$ - $1.0\times10^{18}$ cm$^{-3}$, while DCVB2 crystals fall within the range of $4.6\times10^{17}$ to $1.5\times10^{18}$ cm$^{-3}$. These values are one to two orders of magnitude lower than those of CVB-grown crystals, indicating that DCVB-grown Bi$_2$Se$_3$ crystals contains significantly fewer Se vacancies than their CVB-grown counterparts.

The measurements of temperature-dependent in-plane resistivity further support this conclusion. DCVB1 and DCVB2 crystals with carrier densities near the lower end of the range presented in Table 1 exhibit significantly higher resistivity than CVB-grown crystals. Figure 2(a) shows the temperature-dependent resistivity data for two DCVB-grown crystals with carrier densities of $1.9\times10^{17}$ cm$^{-3}$ (DCVB1) and $4.6\times10^{17}$ cm$^{-3}$ (DCVB2). Across the entire measured temperature range (6–300 K), the resistivity of the DCVB1 sample is approximately an order of magnitude higher than that of the CVB-grown sample. Additionally, above 100 K, the resistivity of the DCVB2 sample also remains significantly higher. Furthermore, the resistivity data of the DCVB1 sample exhibits a broad peak and a small upturn below 30 K. These features closely resemble those observed in previously reported low-carrier-density Bi$_2$Se$_3$ samples grown using the Se-flux method [18-21], indicating significant reduction in bulk conduction.

| Growth Methods | Carrier densities of our Bi$_2$Se$_3$ samples (cm$^{-3}$) | Reported carrier densities of CVB grown pure Bi$_2$Se$_3$ (cm$^{-3}$) | References |
|---|---|---|---|
| Conventional Vertical Bridgman (CVB) | $1.3\times10^{19}$ – $1.6\times10^{19}$ | $2.0\times10^{18}$ – $5.6\times10^{19}$ | [22-26] |
| DCVB1 | $1.9\times10^{17}$ – $1.0\times10^{18}$ | | This work |
| DCVB2 | $4.6\times10^{17}$ – $1.5\times10^{18}$ | | This work |

Table 1. Carrier density range of Bi$_2$Se$_3$ single crystals grown using the conventional vertical Bridgman (CVB) and double crucible vertical Bridgman (DCVB) methods.

To further characterize electron doping in DCVB-grown crystals, we measured the electronic band structure of several pieces from the DCVB2 batch via ARPES at 40 K with a photon energy of 6 eV. Figure 5c shows the ARPES spectrum for one of the measured crystals, which exhibits the typical Dirac-like dispersion of the V-shaped topological surface state. The Dirac point is located at approximately 0.20 eV below the Fermi level. We observe no signatures of bulk conduction band crossing the Fermi level in this sample, suggesting that it is nearly undoped, with the crystal composition likely close to stoichiometric. However, in two other pieces from the DCVB2 batch, we observed Dirac nodes at 0.26 eV and 0.29 eV (see supplementary Fig. S2). The sample with the Dirac node at 0.26 eV shows a slight trace of the bulk conduction band near $E_F$, while the one with the Dirac node at 0.29 eV exhibits a clear bulk conduction band signature near $E_F$. In contrast, the three CVB-grown samples all show Dirac nodes between 0.30 and 0.33 eV, with significant bulk band contributions near $E_F$ (see Fig. 4d and supplementary Fig. S2). These observations not only explain the significantly lower carrier density in DCVB-grown $Bi_2Se_3$ crystals compared to those grown via CVB, but also suggest that the composition distribution in DCVB2 crystals remains somewhat inhomogeneous. This inhomogeneity, which is not detected by EDXS composition analysis (Fig. 3d-f) due to its limited resolution, is likely caused by insufficient Se flux, as discussed above. This issue could potentially be addressed by increasing the Se flux during growth

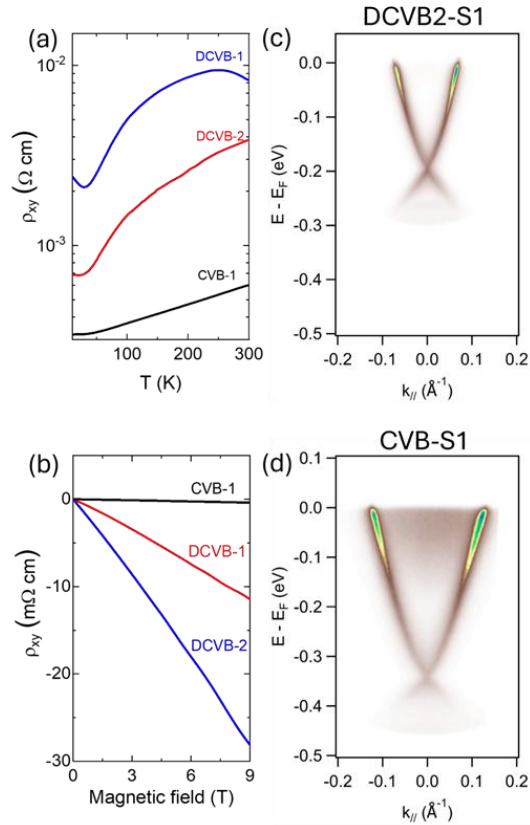

Figure 4. (a) Temperature dependent in-plane resistivity for Bi2Se3 crystals grown using DCVB1, DCVB2 and CVB methods. (d) Field dependent of Hall resistivity for the samples shown in panel (a) measured at 6K. (c) Band dispersion through the Brillouin zone center probed by ARPES at 40 K with the photon energy of 6 eV for a representative DCVB2 crystal. (d) Same as c) for a representative CVB-grown $Bi_2Se_3$ crystal.

## Conclusion

In summary, we present a novel crystal growth technique, the DCVB method, which enables control of crystal composition. The DCVB furnace employs a double crucible geometry, with the upper crucible continuously feeding source material to the growth crucible positioned below. This design enables traveling-solvent growth within a vertical Bridgman furnace. Furthermore, the DCVB furnace employs liquid encapsulation under high pressure to suppress volatile element loss. The combination of these features ensures effective control of the melt composition in the growth crucible, resulting in crystals with better-controlled composition. We applied this technique to grow single crystals of the prototype topological insulator $Bi_2Se_3$ and found that DCVB-grown $Bi_2Se_3$ crystals have carrier densities one to two orders of magnitude lower than those grown by the conventional Bridgman method, suggesting that melt composition control significantly reduces crystal defect (e.g., Se vacancy) density. Additionally, the continuous feeding of material enables the growth of large $Bi_2Se_3$ crystals with low carrier density. This success demonstrates a promising strategy for achieving high-quality, large crystal growth, particularly for metal chalcogenide and pnictide materials with challenging non-congruent melting behaviors

**Method**

**$Bi_2Se_3$ Single Crystal Growth using DCVB**

**DCVB2 growth**: We employed a combination of Bridgman and flux growth methods in the DCVB furnace to grow $Bi_2Se_3$ single crystals, incorporating continuous source material feeding into the growth crucible. Following the previously reported self-flux approach [18], Bi and Se powders with a Bi:Se molar ratio of 34:66 (~2:3.88, corresponding to a Se-rich composition) were thoroughly mixed and loaded into a tip-shaped alumina crucible (inner diameter: 17 mm) under an Ar inert atmosphere inside a glove box.

To prevent the vaporization of volatile components (Se) during heating, 4 g of $B_2O_3$ was placed over the Bi-Se mixture, serving as a liquid encapsulant. The alumina crucible containing the source material was then mounted on the bottom shaft inside the DCVB growth chamber (right panel of Fig. 1b), where it functioned as the growth crucible during crystal growth. The upper crucible, used for feeding the source material, consists of an alumina funnel designed to hold the $Bi_2Se_3$ feed rod and is suspended from the upper shaft (Fig. 1b).

To prepare the feeding rod, $Bi_2Se_3$ single crystals were synthesized using the melt growth method. A mixture of Bi and Se powders with a molar ratio of 2:3 was loaded into a quartz tube, which was then sealed under vacuum. The sealed tube was heated to 900°C and gradually cooled to 600°C. After the furnace was turned off, the grown $Bi_2Se_3$ crystals were extracted and ground into fine powder. A total of 16 g of $Bi_2Se_3$ powder was then loaded into a quartz tube with an inner diameter of 7 mm and heated to 800°C to ensure complete melting. Upon cooling, it solidified into a $Bi_2Se_3$ ingot. The quartz crucible containing the solidified ingot was then placed into the supply crucible, serving as the feeding rod for the DCVB growth process.

After loading both the growth and feeding crucibles into the DCVB furnace, the chamber was initially evacuated to a vacuum level of $10^{-4}$ torr and then filled with high-purity Ar gas to a pressure of 10 atm. The input power of the three carbon heaters was carefully regulated to establish the desired temperature profile. The temperature at the center of the middle zone (i.e., the initial position of the growth crucible) was set to 780°C, while the bottom zone was maintained at 610°C, creating a temperature gradient of 21–23°C/cm from the central to the bottom zone (see Supplementary Fig. S3 for the measured temperature profile). The upper heater was set to 800°C, with the feeding crucible positioned above it.

Once the desired temperature profile was reached, the growth crucible was gradually translated from the initial high-temperature position (780°C) to the low-temperature position (610°C) at a controlled rate of 0.3–0.7 mm/h. Specifically, it was first moved from 780°C to 710°C at a rate of 0.7 mm/h. Upon reaching 710°C—the melting point of $Bi_2Se_3$—the translation rate was reduced to 0.3 mm/h. Simultaneously, the feeding crucible began moving downward at approximately 0.2 mm/min. As it approached the melting point of $Bi_2Se_3$, the $Bi_2Se_3$ rod inside the feeding crucible started to melt and drip into the growth crucible, resulting in a detectable mass change in the feeding crucible. Once this mass change was observed, the translation rate of the feeding crucible was adjusted to 0.9 mm/h. Throughout the growth process, molten $Bi_2Se_3$ was continuously supplied from the feeding crucible to the growth crucible, compensating for the material consumption and preventing the melt from becoming increasingly Se-rich as growth progressed. The source material feeding lasted for ~72 hours, with the total amount of 16 g of Bi2Se3 being fed to the growth crucible. We stopped moving the upper crucible once the feeding was completed while the crystal growth was still on-going

As soon as the growth crucible reached the position where the temperature was 610°C, the power of the central and bottom heaters was gradually ramped down to allow the furnace slowly cooling down to room temperature within 150 hours to minimize thermal fluctuation induced defects [17]. Meanwhile, the temperature of the upper zone was set to ramp down to room temperature at a rate of 21.3 °C/h. Finally, the growth crucible was dismantled to extract the synthesized $Bi_2Se_3$ crystal.

**DCVB1 growth:** For DCVB1 growth without feeding, we followed a similar procedure as DCVB2 growth, except that no source material was supplied during the process.

**$Bi_2Se_3$ Single Crystal Growth using conventional vertical Bridgman** (CVB)

$Bi_2Se_3$ single crystals were grown using a two-step conventional Bridgman method. Stoichiometric amounts of Bi and Se, with an excess of 2% Se to compensate for selenium loss, were sealed in a tip-shaped evacuated quartz ampoule. The mixture was first melted in a rocking furnace at 850°C overnight and rocked for 6 hours before cooling to room temperature to ensure homogeneity and prevent bubble formation within the boule. The homogenized melt was then transferred to a three-zone Bridgman furnace, where the homogenization, crystallization, and solidification zones were maintained at 835°C, 625°C, and 500°C, respectively. Directional solidification was performed under a controlled temperature gradient of ~17°C/cm at 710°C with a translation rate of 2 mm/h until the entire boule passed through the crystallization zone. Upon completion of the growth process, the temperature of all zones was gradually reduced to 500°C for a 2-day annealing step before cooling to room temperature over the course of a day. The *c*-axis of the crystals was oriented perpendicular to the growth direction. The resulting crystals were characterized by X-ray diffraction, transport measurements, and ARPES to assess their electronic properties.

**Characterizations**

Phase identification of the grown crystals was carried out using X-ray diffraction (XRD), while crystal composition and homogeneity were analyzed via X-ray energy dispersive spectroscopy (EDS) and wavelength dispersive spectrometry (WDS) using a CAMECA SX-5 electron microprobe at room temperature.

The electrical transport properties of the $Bi_2Se_3$ crystals were characterized through transport measurements conducted in a Quantum Design Physical Property Measurement System (PPMS). Longitudinal resistivity was measured as a function of temperature over the range of 6–300 K using a standard four-point probe geometry.

Angle resolved photoemission spectroscopy (ARPES) of $Bi_2Se_3$ was performed using 6eV photons from a continuous-wave laser light source (Toptica DLC TA-FHG PRO) with a spot size of 19 μm × 35 μm. Measurements were performed at 40K. We used linear vertically polarized light in combination with a horizontal ARPES analyzer slit. Samples were cleaved below 50K at a pressure of $1 \times 10^{-11}$ torr.

**Acknowledgments:**

The DCVB growth and characterization efforts for $Bi_2Se_3$ were primarily supported by the U.S. Department of Energy under grant DE-SC0019068. The establishment of the DCVB furnace, along with the CVB growth of $Bi_2Se_3$, was supported by the Pennsylvania State University Two-Dimensional Crystal Consortium–Materials Innovation Platform (2DCC-MIP), which is funded by NSF Cooperative Agreement No. DMR-2039351. H.P. and J. O.-G. are supported by the U.S. Department of Energy, Office of Science, Office of Basic Energy Sciences, Materials Sciences and Engineering Division, under Award Number DE-SC0024135.